\documentstyle[12pt,cite]{article}  

\catcode `\@=11
\@addtoreset{equation}{section}

\catcode `\@=12

\title{ 
{}\bf Precanonical Quantum Gravity: 
quantization without the space-time 
decomposition\thanks{\sl PACS classification: {0460, 0420F, 0370, 1110E}}
\thanks{
{\sl Keywords}: Quantum gravity, Hamiltonian formalism, 
quantization (precanonical), 
De Donder-Weyl theory, 
Poisson-Gerstenhaber brackets, 
Clifford algebra, 
covariant Schr\"odinger equation.  }
{}\thanks{to appear in {\em Int. J. Theor. Phys.} {\bf 40} (2001)}}

\author{ 
Igor V. Kanatchikov\thanks{E-mail: 
{\tt kai@tpi.uni-jena.de, ikanat@ippt.gov.pl }} {}\thanks{On leave 
from Tallinn Technical University, 19086 Tallinn, Estonia. } \\
Theoretisch-Physikalisches Institut, \\
Friedrich-Schiller-Universit\"at Jena, \\
Max-Wien-Platz 1, D-07743 Jena, Germany  \\  
\\ 
Laboratory of Analytical Mechanics and Field Theory, \\
 Institute of Fundamental Technological Research, \\
 Polish Academy of Sciences, \\
  \'Swi\c etokrzyska 21,  
Warsaw  00-049, Poland.  
}

\date{} 

\begin{document} 

\maketitle 


 \vspace*{-119mm}\vspace*{-2mm} 

\hbox to 6.25truein{
\hfil \hbox to 0 truecm{\hss 
{\sf  gr-qc/0012074} { }} \vspace*{0.5mm}} 
\hbox to 6.2truein{
\hfil \hbox to 0 truecm{\hss 
{\sf  FSU TPI 14/99} { }}\vspace*{-0.5mm}} 
\hbox to 6.2truein{
\hfil \hbox to 0 truecm{\hss 
{{\em  revised } Dec. 2000} { }}\vspace*{0.0mm}} 

  
\vspace*{95mm} \vspace*{16mm} 






\begin{abstract} 
\normalsize 
A nonpertubative approach to quantum gravity using 
{\em precanonical} field quantization originating from 
the covariant De Donder-Weyl Hamiltonian formulation 
which treats 
space and time variables on equal footing is presented. 
A generally covariant ``multi-temporal'' 
generalized Schr\"odinger equation 
on the finite dimensional space 
of metric and space-time variables is obtained.  
An important ingredient of the formulation is 
the ``bootstrap condition''   
which introduces a classical space-time geometry 
as an approximate concept emerging 
as the quantum average in a self-consistent 
with the underlying quantum dynamics manner.  
An independence of the theory from an arbitrarily fixed 
background is ensured in this way. The prospects 
and unsolved problems of precanonical quantization of gravity 
are outlined.



\end{abstract} 

 
\newpage 


\newcommand{\beq}{\begin{equation}}
\newcommand{\eeq}{\end{equation}}
\newcommand{\beqa}{\begin{eqnarray}}
\newcommand{\eeqa}{\end{eqnarray}}
\newcommand{\nn}{\nonumber}

\newcommand{\half}{\frac{1}{2}}

\newcommand{\xt}{\tilde{X}}

\newcommand{\uind}[2]{^{#1_1 \, ... \, #1_{#2}} }
\newcommand{\lind}[2]{_{#1_1 \, ... \, #1_{#2}} }
\newcommand{\com}[2]{[#1,#2]_{-}} 
\newcommand{\acom}[2]{[#1,#2]_{+}} 
\newcommand{\compm}[2]{[#1,#2]_{\pm}}

\newcommand{\lie}[1]{\pounds_{#1}}
\newcommand{\co}{\circ}
\newcommand{\sgn}[1]{(-1)^{#1}}
\newcommand{\lbr}[2]{ [ \hspace*{-1.5pt} [ #1 , #2 ] \hspace*{-1.5pt} ] }
\newcommand{\lbrpm}[2]{ [ \hspace*{-1.5pt} [ #1 , #2 ] \hspace*{-1.5pt}
 ]_{\pm} }
\newcommand{\lbrp}[2]{ [ \hspace*{-1.5pt} [ #1 , #2 ] \hspace*{-1.5pt} ]_+ }
\newcommand{\lbrm}[2]{ [ \hspace*{-1.5pt} [ #1 , #2 ] \hspace*{-1.5pt} ]_- }
\newcommand{\pbr}[2]{ \{ \hspace*{-2.2pt} [ #1 , #2 ] \hspace*{-2.55pt} \} }
\newcommand{\we}{\wedge}
\newcommand{\dv}{d^V}
\newcommand{\nbrpq}[2]{\nbr{\xxi{#1}{1}}{\xxi{#2}{2}}}
\newcommand{\lieni}[2]{$\pounds$${}_{\stackrel{#1}{X}_{#2}}$  }

\newcommand{\rbox}[2]{\raisebox{#1}{#2}}
\newcommand{\xx}[1]{\raisebox{1pt}{$\stackrel{#1}{X}$}}
\newcommand{\xxi}[2]{\raisebox{1pt}{$\stackrel{#1}{X}$$_{#2}$}}
\newcommand{\ff}[1]{\raisebox{1pt}{$\stackrel{#1}{F}$}}
\newcommand{\dd}[1]{\raisebox{1pt}{$\stackrel{#1}{D}$}}
\newcommand{\nbr}[2]{{\bf[}#1 , #2{\bf ]}}
\newcommand{\der}{\partial}
\newcommand{\oo}{$\Omega$}
\newcommand{\Om}{\Omega}
\newcommand{\om}{\omega}
\newcommand{\eps}{\epsilon}
\newcommand{\si}{\sigma}
\newcommand{\Lm}{\bigwedge^*}

\newcommand{\inn}{\hspace*{2pt}\raisebox{-1pt}{\rule{6pt}{.3pt}\hspace*
{0pt}\rule{.3pt}{8pt}\hspace*{3pt}}}
\newcommand{\sro}{Schr\"{o}dinger\ }
\newcommand{\bm}{\boldmath}
\newcommand{\vol}{\omega}

\newcommand{\bd}{\mbox{\bf d}}
\newcommand{\bder}{\mbox{\bm $\der$}}
\newcommand{\bI}{\mbox{\bm $I$}}

\newcommand{\be}{\beta} 
\newcommand{\ga}{\gamma} 
\newcommand{\de}{\delta} 
\newcommand{\Ga}{\Gamma} 
\newcommand{\gmu}{\gamma^\mu}
\newcommand{\gnu}{\gamma^\nu}
\newcommand{\ka}{\kappa}
\newcommand{\hka}{\hbar \kappa}
\newcommand{\al}{\alpha}
\newcommand{\lapl}{\bigtriangleup}
\newcommand{\psib}{\overline{\psi}}
\newcommand{\Psib}{\overline{\Psi}}
\newcommand{\derts}{\stackrel{\leftrightarrow}{\der}}
\newcommand{\what}[1]{\widehat{#1}}

\newcommand{\bx}{{\bf x}}
\newcommand{\bk}{{\bf k}}
\newcommand{\bq}{{\bf q}}

\newcommand{\omk}{\omega_{\bf k}}



\large 

\section{Introduction }

The goal of contemporary efforts in developing the quantum theory 
of gravity (for a recent review see, e.g., \cite{isham,rovelli}) 
is to complete the synthesis of quantum theory and 
general relativity. This could be achieved either by developing 
a new ``quantum-general-relativistic'' framework in physics 
or by incorporating 
general relativity into a 
unifying quantum theory of all 
interactions. Two 
aspects of classical general relativity, 
which  is at the same time  the  theory of 
space-time 
 and the  theory of 
the gravitational interaction, 
are being stressed then. 
 %
 %
Accordingly, quantization of gravity 
can be viewed either pragmatically, as 
a construction of the quantum field theory of gravity 
(see, e.g., \cite{donoghue,reuter}), or more conceptually, as 
a construction of the quantum theory of space-time.   
Both aspects are of course intimately related to each other.  

By its very nature the program of {\em quantization} of gravity 
is an attempt 
 to apply 
 the  principles 
 of quantum theory, as we understand them at the present,  
to general relativity.  
In fact, it is (necessarily) understood even in 
 a 
more narrow sense 
of imposing the presently known  form of quantization rules,  
which proved to be successful for other fields 
evolving on a fixed 
space-time background, 
to general relativity with its diffeomorphism covariance and 
a dynamical space-time geometry.  
It is  usually assumed that  
the currently practised form of 
quantum (field) theory  
is applicable  to  general relativity without substantial 
modifications (c. f. ``generalized quantum mechanics'' of 
Hartle e.a. \cite{hartle}  which is constructed to be better 
suited to the context of quantum gravity). 
      
 \newcommand{\nabroski}{ 
instead of trying to properly adapt it  to take the peculiarities 
of general relativity into account 
(c.f., however, a few remarkable exceptions .......... 
recent efforts on quantization of diffeomorphism invariant 
theories \cite{ashtekar-ea}  

Quantum Field Theory of Geometry 
Abhay Ashtekar, Jerzy Lewandowski    hep-th/9603083 
Quantization of diffeomorphism invariant theories of connections 
with local degrees of freedom
Abhay Ashtekar, Jerzy Lewandowski, Donald Marolf, Jose Mourao, Thomas Thiemann 
gr-qc/9504018
J.Math.Phys. 36 (1995) 6456-6493
Osterwalder-Schrader Reconstruction and Diffeomorphism Invariance
A. Ashtekar, D. Marolf, J. Mourão, T. Thiemann
quant-ph/9904094

and attempts to modify quantum theory in the form of 
a generalized sum-over-histories quantum mechanics 
\cite{hartle})   

Quantum Mechanics at the Planck Scale 
James B. Hartle  gr-qc/9508023

Spacetime Quantum Mechanics and the Quantum Mechanics of Spacetime, 
J. B. Hartle, 
gr-qc/9304006 

The Spacetime Approach to Quantum Mechanics gr-qc/9210004 
The Quantum Mechanics of Closed Systems gr-qc/9210006

In fact, it is usually understood even in more narrow sense of 
applying the rules of quantization, which proved to be successful 
for other fields, to general relativity,  
without significantly altering the presently known formalism 
to  make it more appropriate to the context of general relativity 
with its 
 } 

 However, it is well known  that the 
 existing attempts to { quantize} gravity 
  are confronted by 
both 
 the problematic mathematical meaning of the involved 
 constructions 
and 
the 
conceptual questions  
originating in 
difficulties of reconciling 
the fundamental principles of quantum theory 
with those of general relativity 
(see, e.g., \cite{isham} for a review 
and further references). 
  In particular, 
a 
 distinct role of the time dimension 
in 
 the 
probabilistic interpretation of quantum theory 
and in 
 the 
formulation of  quantum evolution laws 
seems to contradistinguish from the equal rights status 
of space-time dimensions in the theory of relativity.     
 The quintessential manifestation of this type of difficulties 
is known, somewhat loosely,  as ``the problem of time'' 
(see, e.g., \cite{isham,time} for a review). 
 Besides, 
the commonly adopted procedure of 
{\em canonical quantization} 
  is 
preceded by the Hamiltonian 
formulation which requires 
 a singling out of 
a 
time parameter 
and  seems to be  too 
  tied 
to the classically inspired idea of 
evolution in time from a given Cauchy data. 
 Technically this procedure implies 
a global hyperbolicity of 
space-time which 
  seems to be 
 a rather  unnatural  topological restriction   
for  
the 
expected quantum fluctuating 
``space-time foam''  of quantum gravity.


The difficulties mentioned above are likely to 
indicate that the applicability of the 
conventional Hamiltonian methods 
in quantum theory of gravity can be rather limited.  
However, those difficulties
 could be partially overcome, 
 or at least seen from another perspective,  
if one would have in our disposal 
a quantization procedure in field theory 
which does not so sensibly depend on 
space-time decomposition i.e. on 
 the singling out of  
 a  time parameter. 
 
 One could argue that the path integral approach already 
embodies the idea. However, the particular path integral ansatz of 
Hawking's  Euclidean quantum gravity \cite{hawking} 
  is in fact 
merely a symbolic solution to the Wheeler-DeWitt equation 
  whose derivation  
{\em is} substantially based on 
space-time decomposition. 
Moreover, the interpretation of this ansatz 
  refers to  
 spatial 
 3-geometries (in four dimensions). 
 Besides, 
the usual path integral expression 
of the generating functional in field theory 
incorporates  
 only 
the {\em time ordered } Green functions  
hence, 
an implicit reference to a distinct time parameter.   
    
In fact, what %
we need  instead is 
 a version of canonical quantization without 
 a 
  distinct 
role of 
time dimension 
and, therefore,    independent of the 
picture of fields 
as infinite dimensional 
systems evolving 
in time from the initial Cauchy data 
given on a space-like hypersurface.  
Clearly, 
such a ``timeless'' procedure of quantization, if exists,   
should have to 
be based on  an analogue of 
Hamiltonian 
formalism in which space and time dimensions 
 are 
treated on equal footing.

Fortunately, although this 
 seems to be not commonly  known in theoretical physics,   
the   
Ham\-ilton\-ian-like 
 formulations 
of the field equations 
which could be appropriate for 
 a 
 ``timeless'' version of canonical formalism 
 have been known 
in the calculus of variations 
already at least since the thirties. 
In the simplest version of those formulations,   
the so-called De Donder-Weyl (DW) theory 
 \cite{dw,dickey},  
the Euler-Lagrange field equations 
assume the following 
form of {\em DW Hamiltonian equations}  
\beq
\der_\mu y^a  = \frac{\der H}{\der p^\mu_a}, 
\quad 
\der_\mu  p^\mu_a=- \frac{\der H}{\der y^a } , 
\eeq
where 
$y^a$ denote  field variables, 
$p_a^\mu:=\der L / \der (\der_\mu y^a)$ 
are 
  what we call 
{\em polymomenta},   
$H:=\der_\mu y^a p^\mu_a -L$ is a 
function of $(y^a,p^\mu_a,x^\nu)$ 
called 
the {\em DW Hamiltonian function},   
and  
$L=L(y^a,\der_\mu y^a,x^\nu)$ 
is a Lagrangian density.

Obviously, the above form of the field equations can be viewed as 
a ``multi-temporal'' or multi-parameter generalization 
of Hamilton's canonical equations 
from  mechanics to field theory   
 in which the analogue of the configuration space is 
a 
finite dimensional
space of field and space-time  variables 
$(y^a, x^\nu)$  
and the analogue of the extended phase space is 
a 
finite dimensional (extended) {\em polymomentum phase space} 
 of 
the 
variables $(y^a, p_a^\mu,x^\nu)$. 
In (1.1) fields are described 
essentially as a sort of  
 multi-parameter generalized (DW) Hamiltonian systems 
rather than 
as 
infinite dimensional mechanical 
systems,  
as in the standard Hamiltonian formalism. 
  In doing so,  
the DW Hamiltonian function $H$, which 
thus far does not appear to have any evident 
physical interpretation, 
in a sense controls  
 the 
space-time {\em variations}   
of  fields,     
 as 
specified by 
equations (1.1),       
rather than their time {\em evolution}. 
The latter, however, is  implicit in (1.1)  in the case 
of hyperbolic  field equations for which 
the Cauchy problem can  be posed.  

An intriguing feature of the framework under consideration 
is that in spite of the 
finite dimensionality  of the 
polymomentum  phase space it is 
capable to 
embrace the dynamics of fields 
which  usually are 
viewed as infinite dimensional Hamiltonian systems. 
 From the equivalence of (1.1) 
to the Euler-Lagrange field equations,  
which is only restricted 
by the regularity of the DW Legendre transform  
$y^a_\nu \rightarrow p_a^\nu, 
L \rightarrow H$, 
it is obvious that no field degrees of freedom 
 are lost when transforming to the DW formulation.  
 In fact, instead of the standard notion of 
a degree of freedom 
 per 
 space point, which 
 originates in the conventional Hamiltonian treatment, 
in the present multi-parameter Hamiltonian   
description it is the (finite) number of the components of 
the field, 
which is important.  
The label of the conventional field degree of freedom, 
the space coordinate {\bf x}, 
goes over to the space-time multi-parameter 
{$x^\mu = ({\bf x},t)$};  that is  
the usual ``infinite-dimensionality'' of field theory 
in the present formulation  
 is equivalently  
 accounted for   
 in the form of 
 ``multi-parametricity.''   
 The same  also 
 applies to field quantization  based on DW theory, 
which is to be described below.

Note that there exists an analogue of the Hamilton-Jacobi 
theory corresponding to the  DW Hamiltonian 
equations (1.1). The DW Hamilton-Jacobi equation 
  \cite{dw,vonrieth} 
is formulated 
for $n$ ($n$ = the number of space-time dimensions) functions 
$S^\mu=S^\mu(y^a,x^\mu)$:   
\beq
\der_\mu S^\mu + H (x^\mu,y^a,p_a^\mu=\der S^\mu/\der y^a)=0 .     
\eeq 
It naturally leads to the question as to 
which formulation of quantum field theory 
could yield 
this field theoretic 
Hamilton-Jacobi equation in the classical limit.  
The scheme of field quantization which is 
outlined in Section 2.2 is a possible answer 
to the question 
and underlies the present approach to quantization of gravity. 

It should be mentioned that the DW formulation 
 is a particular case of more 
general {\em Lepagean} \cite{lepage} 
Hamiltonian-like 
theories 
 for 
fields 
 (known in the calculus of variations of multiple integrals \cite{giaq})  
which differ by 
 the 
definitions of polymomenta and  
the analogues of Hamilton's canonical function 
$H$ (both following 
essentially from different choices of the 
Lepagean equivalents of 
the Poincar\'e-Cartan form; for further details and references 
see \cite{dedecker,krupka,gotay-ext}   
and the reviews quoted in \cite{dw}).  
All 
theories of this type 
 treat space and time variables on equal footing and 
are finite dimensional in the sense that the 
corresponding analogues 
 of the configuration 
and the polymomentum   phase space are finite 
dimensional.  They all reduce to the Hamiltonian 
formalism of mechanics 
at $n=1$. 

Moreover, 
all these formulations are, in a sense, 
intermediate between the Lagrangian formulation and 
the canonical Hamiltonian formulation: they still keep space-time 
variables indistinguishable but already possess the 
essential features of the Hamiltonian-like description 
being based on the first order form of the field equations 
and a Legendre transform.   
 Besides, 
 there are intimate relations, 
not fully studied as yet, between the 
structures of the canonical Hamiltonian formalism 
and the structures of the  
Lepagean formulations \cite{gotaymulti1,gimmsy,helein} 
which point to the latter as a natural intermediate step 
when formulating the field theories canonically proceeding from 
the Lagrangian level.   
For this reason,  
 henceforth  we refer to the finite dimensional covariant Hamiltonian-like 
formulations based on different Lepagean theories as 
``{\em precanonical\,}''. 
Further justification of the term can be found in 
 Section 4.   
The term ``{\em precanonical quantization}'' to be used throughout 
means, in most general sense, a quantization based on the 
Hamiltonian-like structures of a Lepagean theory. 
In this paper, however, we deal only with a particular Lepagean 
theory: the DW formulation and the corresponding quantization. 
Thus the term precanonical quantization will be used rather 
in this limited sense.


Let us note also 
that precanonical formulations 
typically have different regularity conditions than the canonical 
Hamiltonian formalism. For example, the DW formulation (1.1) requires 
that $\det || \der^2 L / \der_\mu y^a \der_\nu y^b || \neq 0$. 
This condition is obviously 
different from the regularity condition of the canonical formalism: 
$\det || \der^2 L / \der_t y^a \der_t y^b || \neq 0$.  
As a result, the ``constraints'', 
understood as obstacles to the corresponding 
generalized Legendre transforms   
 $\der_\mu y^a \rightarrow p^\mu_a$,  
have 
 a 
quite different structure from the standard canonical formalism.  
 In fact, the singular theories from the point of view of the 
 canonical formalism can be regular from the precanonical point of view 
(as e.g. the Nambu-Goto string \cite{romp98}) or vice verse 
(as e.g. the Dirac spinor field \cite{vonrieth}).  
This opens an yet unexplored 
possibility 
of avoiding the constraints analysis when quantizing 
within the precanonical framework 
by choosing for a given theory an appropriate non-singular 
Lepagean Legendre transformation. In fact, this possibility is 
exploited below, in Section 3.2, 
when quantizing general relativity without 
any mentioning of constraints

The idea of using the 
DW Hamiltonian formulation 
for field quantization 
dates back to Born (1934) and Weyl (1934) 
\cite{bw}.  
However it has 
not received much attention 
since then 
(see, however, \cite{attempts}). 
Obviously,  one of the reasons is that 
quantization needs more than just an existence of 
a Hamiltonian-like formulation 
of the field equations: additional structures,   
such as the Poisson bracket (for  canonical or  deformation 
quantization),  the symplectic structure 
(for  geometric quantization),   
and 
a 
Poisson bracket formulation of the 
field equations (in order to 
formulate or postulate 
the 
quantum dynamical law) are 
necessary. 

Unfortunately, 
in spite of a number of  earlier 
attempts \cite{oldbrackets} and 
 the progress in understanding 
the relevant aspects of the geometry of classical field theory, 
such as those related to the notion of the 
\mbox{(Hamil\-ton-{})}Poincar\'e-Cartan  
 (or multisymplectic) form 
\cite{sternberg,crampin,gimmsy} 
 and  G\"unther's polysymplectic form \cite{guenther},  
 a construction which could be suitable 
 as a starting point of  quantization   
has been lacking.  
It is only recently that a  
proper Poisson bracket operation, 
which is defined on differential forms representing 
  the dynamical variables and leads to a
Poisson-Gerstenhaber algerba structure,   
has been found within the DW theory 
in \cite{ik93,bial94,romp98,bial96,goslar96}   
(see also \cite{helein,paufler,roemer} for recent generalizations). 
This progress has been accompanied and followed 
by further developments in 
``multisymplectic'' generalizations of the symplectic geometry 
aimed at applications in field theory and the calculus of 
variations \cite{sardan,deleon,cantrijn,norris,betounes,olver}  
and in other geometric aspects of the Lagrangian and Hamiltonian 
formalism in field theory \cite{echev1,rr1,rr2,rr3,rr4,rr5,hrabak,ik-dkp} 
which to a great extent  are been so far basically 
ignored by the wider  mathematical physics community.

The elements of field quantization  
based on the aforementioned Poisson-Gerstenhaber 
brackets on differential forms have been  discussed in  
 \cite{bial94,qs96,bial97,lodz98} 
and will be briefly summarized in Section 2.2.     
 Unfortunately, many fundamental aspects of the corresponding 
{\em precanonical} approach to  field quantization, 
as we suggest to call it,    
 so far remain  poorly understood   
(see \cite{qs96,lodz98} and Section 4 for a discussion)  
and require a further analysis.  
This particularly concerns an interplay with 
the standard formalism and notions of quantum field theory 
(see \cite{ik-schr} for a recent progress).  
 Nevertheless,   
 the already elaborated part of the theory points to 
 intriguing features  and as yet unexplored potential  
 which, hopefully,  are capable to make the precanonical 
approach a useful 
complement to the presently available concepts 
and techniques of quantum field theory.  

The purpose of the present paper 
is to  apply  the  precanonical approach 
to field quantization,  as we understand it now, 
to the problem 
of quantization of general relativity (see \cite{ik-gr,ik-gr2,ik-gr3,ik-gr4} 
for earlier reports).   
We hope that this application 
can shed new light on the problems of quantum gravity 
and  can be useful also for better understanding of the 
precanonical approach itself. 

We proceed as follows: 
first, in Section 2, we summarize 
basic elements of 
precanonical formalism and   
quantization 
based on the DW theory 
and then, in Section 3, 
apply  
this framework  
to general relativity. 
Discussion and 
concluding remarks 
are presented in Section 4.

\section{Precanonical formalism and quantization 
based on 
DW theory }

In this section we briefly summarize  basic  elements 
of 
precanonical formalism based on 
the DW theory and then  outline 
the corresponding precanonical field quantization scheme.

\subsection{Classical theory }  
The mathematical structures underlying 
 the DW form of the field equations, eq. (1.1),  
 have been studied in our previous papers 
\cite{romp98,bial96,goslar96} to which we refer for more details.

The 
analogue of the  
Poisson bracket in the DW formulation 
is deduced from the object, 
called the {\em polysymplectic form, } 
which in local coordinates 
can be written in the form\footnote{Strictly speaking 
this object is understood 
as the equivalence class of forms modulo  the forms of the 
horizontal degree $n$, see \cite{romp98} for more details. 
Henceforth we denote  
$ \omega := dx^1\we ... \we dx^n , \quad 
\omega_\mu := \der_\mu\inn \omega = 
(-1)^{\mu -1}dx^1 \we ... \what{dx^\mu} ... \we dx^n $, 
$dx\uind{\mu}{p} :=  
dx ^{\mu_1}\we ...\we dx^{\mu_p} $,  
and $\{z^M \} :=  \{y^a,p_a^\mu,x^\mu \} $.} 
 \[ 
\Omega = - dy^a \we dp_a^\mu \we \omega_\mu 
 \] 
and is viewed as a field theoretic generalization
 of the symplectic form within the DW formulation. 
 Note that if  
$\Sigma$,  
$\Sigma\!:\!(y^a\!=\!y^a_{}(\bx), \, t\!=\!t_{})$, 
denotes the Cauchy data surface 
in the covariant configuration space $(y^a,x^\mu)$  
the standard symplectic form in field theory, $\omega_S$,  
 can be expressed as 
 the integral over $\Sigma$  of  the restriction 
of $\Omega$ to $\Sigma$, $\Omega|_\Sigma$ 
\cite{gotaymulti1,gimmsy},  
 i.e.   
$$\omega_S = \int_\Sigma \Omega|_\Sigma.$$    
 
The polysymplectic form $\Omega$ associates   
horizontal $p$-forms $\ff{p}$, 
$\ff{p}: 
= \frac{1}{p!}  
F\lind{\mu}{p}(z^M)
 dx\uind{\mu}{p} 
$, { }  
($p=0,1,...,n$), 
with  $(n-p)$-multivectors 
(or more general algebraic operators 
 of degree $-(n-p)$ 
on the exterior algebra),  
 $\xx{n-p}$, by the relation:  
\beq 
\xx{n-p} \!\inn \Omega 
= d \ff{p} ,  
\eeq   
where $\inn$ denotes the contraction of a multivector 
with a form.  Then the graded Poisson brackets  
of horizontal forms representing the  
dynamical variables 
is given by 
\beq
\pbr{\ff{p}{}_1}{\ff{q}{}_2} := (-)^{n-p} \xx{n-p}{}_1 \inn d \ff{q}{}_2.   
\eeq  
Hence the bracket of a $p$-form with a $q$-form is a form of degree 
$(p+q-n+1)$, where $n$ is the space-time dimension. 
Note that, as a consequence, the subspace of forms of degree $(n-1)$  
is closed with respect to the bracket, as well as the subspace of 
forms of degree $0$ and $(n-1)$. 

The above construction leads to 
a hierarchy of algebraic structures which are graded 
generalizations of 
 the 
Poisson algebra in mechanics \cite{romp98,bial96,goslar96}. 
Specifically, 
on a small subspace 
of the so-called Hamiltonian forms 
(i.e. those which can be mapped   
by  relation (2.1) to multivectors)   
one obtain the structure 
of a so-called Gerstenhaber algebra \cite{gerst}. 

Let us recall, that the latter is a triple 
${\cal G}=({{\cal A},\bullet,\pbr{\ }{\,}})$, 
where ${\cal A}$ is a graded commutative associative 
algebra with the product operation $\bullet$ and 
$\pbr{\ }{\, }$ is a graded Lie bracket 
which fulfils the graded Leibniz rule with respect 
to the product $\bullet$, with the degree of an element 
$a$ of ${{\cal A}}$ with respect to the bracket 
operation, $bdeg(a)$, 
and the degree of $a$ with respect to the 
product $\bullet$, $pdeg(a)$, related as $bdeg(a)=pdeg(a)+1$.  
In our case the Lie bracket is the bracket operation 
defined in (2.2) which is also closely related 
to the Schouten-Nijenhuis bracket of multivector fields 
(the latter is related to our bracket in the similar  
way as the Lie bracket of vector fields is related to the Poisson bracket).  
Correspondingly,  the graded commutative multiplication 
 $\bullet$ is what we call the ``co-exterior product'' 
 \[ 
F\bullet G:=*^{-1}(*F\we*G) 
\] 
($*$ is the Hodge duality operator),    
with respect to which the space of Hamiltonian 
forms is stable \cite{bial96,goslar96,paufler}.    
 
Note that more general (``non-Hamiltonian'') forms 
give rise to a 
non-commuta\-tive 
(in the sense of Loday's ``Leibniz algebras''  
\cite{loday})    
higher-order (in the sense of a higher-order 
analogue of the graded Leibniz rule 
replacing the standard Leibniz rule 
in the 
definition)  
generalization of a Gerstenhaber algebra 
\cite{bial96,goslar96}.  

The bracket defined in (2.2) 
enables  
us 
to identify 
the 
pairs of 
``precanonically conjugate'' variables 
and to represent the DW Hamiltonian equations 
in (generalized) Poisson bracket formulation.  
In fact, the appropriate notion of 
precanonically conjugate variables in the present 
context is suggested by considering 
the brackets of horizontal forms 
of the kind 
$y^a dx^{\mu_1}\we ... \we dx^{\mu_p}$ 
and $p_a^\mu \der_\mu \inn\der_{\mu_1}\inn ... 
\der_{\mu_q}\inn \omega ,$  
with $p\geq q$. 
In particular, 
in the Lie subalgebra of 
 Hamiltonian forms of degree 
$0$ and $(n-1)$ the non-vanishing 
 ``precanonical'' brackets 
take the form \cite{romp98}
\beqa 
\pbr{p_a^\mu\omega_\mu}{y^b}
&=& 
\delta^b_a ,  
 \nn \\
\pbr{p_a^\mu\omega_\mu}{y^b\omega_\nu}
 &=&
\delta^b_a\omega_\nu,  
 \\ 
\pbr{p_a^\mu}{y^b\omega_\nu}
 &=&
\delta^b_a\delta^\mu_\nu   .  
\nn 
\eeqa 
This brackets obviously reduce to the canonical Poisson bracket 
in mechanics, $\{p^a,q_b \}=\delta^a_b$,  at $n=1$. 
Hence, the 
 pairs of variables entering  the brackets (2.2), (2.3) 
can be viewed as precanonically conjugate 
 with respect to the graded Poisson 
bracket (2.2). 
Note that the brackets (2.3) do not involve any dependence 
on space and time variables. Therefore, they  
can be viewed as ``equal-point'' brackets, 
as opposite to the usual ``equal-time''  Poisson 
brackets in field theory. 

 By considering the brackets of 
precanonical variables entering (2.3) 
with $H$ or $H\omega$  
we can write the DW Hamiltonian field equations 
(1.1) in Poisson bracket formulation:  
for example,    
 \beqa 
\bd(y^a\omega_\mu) 
&=&  \pbr{H\omega}{y^a\omega_\mu} = \frac{\der H }{\der p^\mu_a} \,\omega , 
\nn \\ 
\bd(p^\mu_a \omega_\mu) 
&=& \pbr{H\omega}{p^\mu_a \omega_\mu} = \frac{\der H }{\der y^a}\,\omega ,  
\eeqa 
where $\bd$ is the total exterior differential 
such that e.g.  
$\bd y = \der_\mu y(x) dx^\mu $. 
The 
DW Hamiltonian equations written in the form (2.4) 
point to the fact that 
 the 
 type of the space-time variations  
which are controlled by $H$ 
is 
related to the operation of the exterior 
differentiation. This generalizes to the 
present formulation of field theory the 
familiar statement 
in the analytical mechanics 
that  
Hamilton's canonical function generates 
 the 
time evolution. 
Note that this  observation largely 
underlies our hypothesis (2.6)   regarding 
the form of a  generalized Schr\"odinger equation 
within 
the 
precanonical quantization approach \cite{bial94,qs96,bial97}.

\subsection{Precanonical quantization }   

Quantization of the Gerstenhaber algebra 
${\cal G}$ or its above-mentioned generalizations 
 would be a difficult mathematical problem. 
One may even  doubt that the current notions of quantization or deformation 
are general enough to treat the problem \cite{flato}. 
This is due to the fact that $bdeg(a)\neq pdeg(a)$ for $a \in {\cal G}$. 
This is  only recently that a progress has been made 
along the lines of geometric quantization of 
the Poisson-Gerstenhaber 
brackets (2.2) \cite{mg9a} which 
suggests that the difficulty can be solved by admitting the 
operators to be nonhomogeneous in degree, at least on 
the level of prequantization.

Fortunately, in physics we usually do not need to quantize the whole  
Poisson algebra. It is even known to be impossible, in the sense of 
Dirac canonical quantization, as it follows from the 
Groenewold-van Hove ``no-go'' theorem 
\cite{emch,gotay-q}.     
In fact,  quantization of 
a 
small Heisenberg subalgebra of the canonical 
brackets often suffices.  

It seems reasonable, therefore, at least as 
the 
first step,  
 to  quantize a small subalgebra of graded Poisson brackets 
which 
resembles the Heisenberg subalgebra of canonical 
variables.  A natural candidate is the 
subalgebra of precanonical brackets in the 
Lie subalgebra of Hamiltonian forms of degree $0$ and $(n-1)$, 
eqs. (2.3). 	
In fact, the scheme of field quantization 
discussed in \cite{bial94,qs96,bial97}  
is essentially based on  quantization of this 
small subalgebra by the Dirac correspondence rule: 
$[\hat{A},\hat{B}] = i\hbar \pbr{A}{B}.$ 
It leads to  
the following realization of operators corresponding 
to the  quantities involved in (2.3) :   
\beqa  
\widehat{p_a^\mu\omega_\mu}&=& i \hbar 
 \,\der / \der y^a 
 , \nn \\
\hat{p}{}_a^\nu &=&  - i \hbar  \kappa \ga^\nu \
  \der / \der y^a 
  ,  \\  
 \widehat{ \omega}_\nu&=&  - \kappa^{-1} \ga_\nu 
  ,  \nn 
\eeqa 
where $\gamma^\mu$ are 
the 
imaginary units of   
the Clifford algebra 
of the space-time manifold and    
the parameter $\kappa$ of the dimension 
{(length)}$^{-(n-1)}$ is 
required by 
the dimensional consistency of (2.5). 
An identification 
 of $\kappa$ with   
the ultra-violet cutoff 
or 
a fundamental length scale 
 quantity was discussed in \cite{qs96,lodz98}. 
 The realization (2.5) is essentially inspired  by 
the relation between the  Clifford algebra and 
the endomorphisms of the exterior algebra \cite{chevalley}. 
 A crucial assumption 
underlying the proof 
 that the operators in (2.5) fulfil 
 the 
commutators following from (2.3) 
is  that the composition law of operators 
 implies  
 the symmetrized product of $\gamma$-matrices. 

The realization (2.5) 
indicates that quantization of DW formulation, 
viewed as a multi-parameter generalization 
of the standard  Hamiltonian formulation 
with a single time parameter, 
  results in a generalization   of 
 the 
quantum theoretic formalism in which 
 (i) the hypercomplex (Clifford) 
algebra of the underlying space-time manifold 
 replaces 
the algebra of the complex numbers (i.e. the Clifford algebra of 
(0+1)-dimensional ``space-time'') 
in quantum mechanics, 
and (ii) $n$ space-time variables 
 are treated on 
equal footing 
 and 
 generalize 
the one dimensional time parameter. 
In doing so the quantum mechanics is reproduced as a special 
case corresponding to $n=1$. 

This 
 philosophy 
 leads to 
the following (covariant, ``multi-temporal'', hypercomplex) 
generalization of  the Schr\"odinger equation to the 
precanonical framework \cite{qs96,bial97,lodz98}  
\beq
\label{seqcl}
i \hbar \kappa \gamma^\mu \der_\mu \Psi = \what{H} \Psi, 
\eeq
where $\widehat{H}$ is the operator corresponding to the 
DW Hamiltonian function, the constant $\kappa$ 
appears again on dimensional grounds, 
and $\Psi=\Psi(y^a,x^\mu)$ is the 
wave function over 
the covariant configuration space of 
 field and space-time variables. 
 

Equation (2.6) gives rise to the conservation law 
\beq
\der_\mu \int \!dy\, \Psib \gamma^\mu \Psi = 0 
\eeq 
provided $\widehat{H}$ is Hermitian with respect to the 
scalar product 
$\left < \Psi,\Phi\right > = \int \!dy\, \overline{\Psi}\Phi$,   
which is also used for 
calculating the expectation values of operators: 
 \beq 
\langle\,\what{O}\,\rangle  (x) 
 := \int \!dy\, \Psib\what{O}\Psi .  
\eeq

The main argument in favour of 
a generalized Schr\"odinger equation (2.6)  
is that  it satisfies at least two important 
aspects of the correspondence principle   
\cite{bial97,lodz98}: 

(i) it leads, at least 
 in the simplest case of scalar fields, 
to the DW canonical equations (1.1) for the mean values 
of the appropriate operators (the Ehrenfest theorem), 
 e.g.,  
\beqa
\der_\mu \left < \what{p}{}^\mu_a\right > &=&
-\left < \left ({\der H}/{\der y^a}\right )^{op}\right >,  \nn \\  
 \der^\mu \left < ({y^a\om}{}_\mu)^{op} \right > &=& 
\left < \left ({\der H}/{\der p^\mu_a} \; \om_\mu \right )^{op} \right >,  
 \eeqa 
where ${(F)}^{op}$ denotes the operator corresponding to the 
variable  $F$, 
and 

(ii) it reduces to 
the DW Hamilton-Jacobi equation (1.2) 
(with some additional conditions) 
in the classical limit.    

Moreover, it was shown recently that eq. (2.6) 
allows 
us to derive the standard functional differential 
Schr\"odinger equation once a suitable physically 
motivated ansatz relating the Schr\"odinger wave 
functional and the wave function in (2.6) 
is constructed \cite{ik-schr}. 

Some details on the  application of the 
 present precanonical quantization scheme 
to the case of scalar fields can be found in \cite{bial97,lodz98,ik-schr}.  
It should be noted that a capability of eq. (2.6) to reproduce 
in the classical limit 
the field  equations,  
i.e.  
an infinite dimensional 
Hamiltonian system in the conventional sense,   
  implies that despite 
 the generalized Schr\"odinger equation, eq. (2.6), 
is partial differential and  is formulated in terms of 
a finite dimensional analogue of the 
configuration space no 
field degrees of freedom 
are lost in this description. Similarly to the classical level, 
the  customary ``infinite-dimensionality'' goes over into 
a ``multi-parametricity''. Further  details on the interplay 
between precanonical and canonical field quantization 
have been discussed recently in \cite{ik-schr}.


\section{Precanonical quantization of general relativity}

In this Section we first outline 
 a 
curved space-time generalization 
of 
 the precanonical quantization scheme 
presented in  Section 2.2 and then discuss  its further 
 application to quantization of general relativity. 
 The required DW Hamiltonian formulation of general relativity 
is discussed in Section 3.2.1.  
The rest of Section 3.2 is devoted to the derivation of  
a  diffemorphism covariant 
 Dirac-like wave equation for quantum  general relativity.  
 This equation is argued to include a ``bootstrap condition'' 
 which introduces an  
 averaged  self-consistent classical geometry 
 involved in  the  Dirac-like wave  equation 
 ensuring, in this sense,  the independence  
 of the formulation from 
 the choice of an arbitrary background.

\subsection{ Curved space-time generalization } 
To apply the 
precanonical framework 
to general relativity we 
first need to extend 
it 
to curved space-time with the metric 
$g_{\mu\nu}(x)$. 
The extension of the generalized Schr\"odinger 
equation (2.6) to 
curved space-time is  similar to that of the Dirac equation, i.e. 
\beq  
i\hbar\kappa 
\gamma^\mu (x) \nabla_\mu \Psi , 
= \what{H} \Psi 
\eeq
where 
$\what{H} $ is 
an 
operator form of 
the DW Hamiltonian function    
and 
$\nabla_\mu$ is 
 the 
covariant derivative, 
$\nabla_\mu :=  \der_\mu + \theta_\mu (x)$.   
We introduced 
$x$-dependent $\gamma$-matrices 
which fulfil 
\beq
\gamma_\mu(x) \gamma_\nu(x) + \gamma_\nu(x) \gamma_\mu(x)
=2 g_{\mu\nu}(x)     
\eeq 
and can be expressed with the aid  of 
vielbein fields 
$e_\mu^A (x)$,  such that 
\beq
g_{\mu\nu}(x) = 
e_\mu^A (x) e_\nu^B (x) \eta_{AB}, 
\eeq 
and the 
(pseudo-)Euclidian tangent space 
Dirac matrices $\gamma^A$, 
$\gamma^A\gamma^B + \gamma^B\gamma^A 
:= 2 \eta^{AB}$:     
\[
\gamma^\mu (x):= e^\mu_A (x) \gamma^A. 
\]

  If $\Psi$ is a spinor wave function then 
$\nabla_\mu$  is the spinor covariant derivative: 
$\nabla_\mu =  \der_\mu + \theta_\mu$, 
where 
 \[ 
\theta_\mu = \frac{1}{4} \theta_{AB}{}_\mu \gamma^{AB}, 
\quad \gamma^{AB}:= \half (\gamma^A\gamma^B - \gamma^B\gamma^A) 
\]  
 denotes the spin connection with the  components 
given by the usual formula 
\beq
\theta^A{}_{B\mu} = 
 e^A_\al e^\nu_B \Ga^\al{}_{\mu\nu} 
 -e^\nu_B \der_\mu e^A_\nu .
\eeq

For example, interacting scalar fields $\phi^a$ on 
a curved background 
are described by 
the Lagrangian density  
 \[
{\cal L}=  
\sqrt{g} \{ \half \der_\mu \phi^a \der^\mu \phi_a 
 - U(\phi^a) 
- \xi R \phi^2 \} ,  
\]  
where $g:={|\rm det}(g_{\mu\nu})|$.  
 This 
gives rise to the following 
 expressions of 
polymomenta and the DW Hamiltonian density 
 \[ 
p^\mu_a := \frac{\der {\cal L}}{\der(\der_\mu \phi^a)} =  
\sqrt{g}\der^\mu \phi_a ,  
\quad 
\sqrt{g} H =  
\frac{1}{2\sqrt{g}}p^\mu_a p^a_\mu + 
\sqrt{g} \{ U(\phi)
+ \xi R \phi^2  \} 
\]
 for which  the corresponding operators 
  can be found to 
 take the form 
 \beqa
\what{p}{}^\mu_a &=&  
- i\hbar\kappa \sqrt{g} \gamma^{\mu}\frac{\der}{\der \phi^a} , 
  \nn \\ 
\what{{H}} 
&=& -\frac{\hbar^2\kappa^2}{2}
\frac{\der^2}{\der \phi^a \der \phi_a} 
+  
U(\phi) 
+ \xi R \phi^2 
 . 
 \eeqa

\subsection{Precanonical approach to quantum general relativity }  

In the context of general relativity 
the field variables are the 
metric 
$g_{\al\beta}$ 
(or the vielbein $e_A^\mu$) 
components. 
Hence, according to the precanonical scheme,   
the wave function is  
a function 
of space-time and metric (or vielbein) variables, i.e. 
$\Psi= \Psi (x^\mu, g^{\al\beta})$ (or $\Psi= \Psi (x^\mu,e_A^\mu)$).    
To formulate an analogue 
of the Schr\"odinger equation for this wave function we  need 
$\gamma$-matrices which fulfill  
\beq
\gamma^\mu \gamma^\nu + \gamma^\nu \gamma^\mu
= 2 g^{\mu\nu}    
\eeq
 and 
are 
related to the 
(pseudo-)Euclidian $\gamma$-matrices $\gamma^A$  
by 
the vielbein components:
$\gamma^\mu := e_A^\mu \gamma^A$,  
with  $g^{\mu\nu}=: e_A^\mu e_B^\nu\eta^{AB}$. 
Note that,  as opposite to the theory on curved 
background,  the  variables 
$e_A^\mu $, $\gamma^\mu$ and $g^{\mu\nu}$  
do not carry any dependence on space-time 
variables $x$; they are instead viewed as  
  the 
fibre coordinates in the corresponding 
bundles over the space-time. 
The corresponding fields $e_A^\mu (x)$, $\gamma^\mu (x)$ 
and $g^{\mu\nu}(x)$  exist only as classical notions 
and represent 
 the 
sections in 
these bundles.

Now, modelled after (3.1),  
the following (symbolic form of the) 
generalized Schr\"o\-din\-ger equation 
for the wave function of quantized gravity can be 
put forward 
\beq
i\hbar\kappa 
\what{ \mbox{$\hspace*{0.0em}e\hspace*{-0.3em}
\not\mbox{\hspace*{-0.2em}$\nabla$}$}} 
 \Psi = 
\what{{\cal H}\hspace*{-0.0em}} \Psi  ,   
\eeq 
where  $\what{{\cal H}}:= \what{e H}$, 
is the operator form 
of the DW Hamiltonian 
 density  
of gravity, 
an explicit form of which is to be constructed, 
$e:= |{\rm det}(e^A_\mu)|, $ 
and 
$\what{\mbox{$\hspace*{0.0em}\not\mbox{\hspace*{-0.2em}$\nabla$}$}}$  
denotes the quantized Dirac operator in the sense that 
the corresponding connection coefficients are 
replaced by appropriate differential operators  
(c.f., e.g.,  eqs. (3.15), (3.16) and (3.21) below).   
Note also, that 
in the context of quantum gravity it 
seems 
 to be 
very natural to identify the parameter $\kappa$ in (3.7) 
with the Planck scale quantity, i.e. 
$\kappa \sim \ell_{\mbox{\footnotesize Planck}}^{-(n-1)}$.  


If the wave function in (3.7) is  spinor  
then the covariant derivative operator   
$\what{\nabla}{}_\mu $ contains the spin connection which 
on the classical level involves 
 the term with  
the 
space-time derivatives 
of vielbeins  (c. f. eq. (3.4)) 
which cannot be expressed in terms of 
the 
quantities 
of the metric formulation. 
Consequently, the spinor nature of  equation (3.7)  
 seems to necessitate the use of the vielbein formulation of general 
relativity.
However,  
 no  suitable DW formulation of general relativity 
 in  vielbein variables is available  
so far.  
(for a related discussion see also \cite{esposito}).   
The main problem is that the Lagrangian in vielbein formulation  
depends on vielbeins ($4\times 4$ components in $n=4$ dimensions) 
and the spin connection  
($4\times 6$ components) which  involves only the 
antisymmetrised space-time derivatives of vielbeins. 
Hence, the space-time derivatives of vielbeins  
($4\times 4 \times 4$ components) 
cannot be expressed in the desired DW Ha\-mil\-ton\-ian form 
$\der_\mu e^a_\nu = \der H / \der \pi^\nu_a{}^\mu$  
(c.f. eq. (1.1)) for any 
definition of $H$ and 
 polymomenta $\pi^\nu_a{}^\mu$ 
because the latter will 
be constrained to be antisymmetric in indices $\mu$ and $\nu$.  
The similar problem is encountered in DW formulation of 
electrodynamics \cite{romp98}  
due to the irregularity of DW Legendge transform which is a consequence 
of the presence of only the 
antisymmetrised space-time derivatives of four-potentials 
in the Lagrangian.  
The problem, however, can be avoided    
if one  starts from a proper gauge fixed action 
\cite{inprogress}  
(note, that this step can be interpreted also 
as a choice of another Lepagean equivalent of the Lagrangian).  
 In fact, what we need here is a precanonical analogue of 
the 
analysis of 
irregular (in the sense of DW formulation) Lagrangians and the 
corresponding quantization. Unfortunately, 
this part of the theory remains so far  
 to a great extent  
 undeveloped. 
For this reason   
the subsequent 
consideration   
 will be based on 
the metric formulation which does not suffer from the 
above problems 
because  
the Lagrangian depends on the Christoffel 
symbols ($4\times 10$ components) and thus enables us to 
express 
the 
first derivatives of the metric ($4\times 10$ components) 
in DW form (c.f. Section 3.2.1 below).  
 
As we shall see,  the metric formulation also enables us 
to discuss 
 the 
basic ingredients of precanonical quantization 
of gravity. 
In fact, one can argue that the additional degrees of freedom 
of the vielbein gravity, as compared with the metric gravity, 
that is those related to the local orientations of vielbeins,  
are not physical: they have to be gauged away 
by a coordinate gauge condition which has to be imposed  
in the end of the quantization procedure (c. f. eq. (3.24) below).  
This makes the analysis 
based on the metric formulation even 
more justified from the physical point of view.  

\subsubsection{DW formulation of the Einstein equations. }  

A suitable DW-like   
formulation of general relativity in metric variables 
was presented earlier by Ho\v rava \cite{horava}. 
 In this formulation 
 the 
field variables are chosen to be the metric density components 
$ h^{\alpha\beta}:= \sqrt{g}g^{\al\be}$ 
and the polymomenta, $Q^\al_{\beta\gamma}$, are found to be 
represented by  the following combination of the Christoffel symbols 
\beq 
Q^\al_{\beta\gamma} := 
 \frac{1}{8\pi G}( \delta^\alpha_{(\beta}\Gamma^\delta_{\gamma)\delta} 
  - \Gamma^\al_{\beta\gamma}) .   
\eeq 
Respectively, 
the DW Hamiltonian 
{ density}  ${\cal H}:= \sqrt{g} H$   
 assumes 
 the form 
\beq
{\cal H} 
(h^{\alpha\be}, Q^\al_{\beta\gamma}) :=  
 8\pi G\, 
h^{\alpha\ga} \left ( 
Q^\de_{\al\be } Q^\be_{\ga\de }+ 
\frac{1}{1-n}\, Q^\be_{\al\be }Q^\de_{\ga\de } \right )
+(n-2) \Lambda \sqrt{g}  
\eeq  
which is 
essentially the 
truncated Lagrangian 
density of general relativity 
 (with the opposite sign of the cosmological term)    
 written in terms 
of variables $h^{\alpha\be}$ and 
$Q^\al_{\beta\gamma}$. 

Using these variables the Einstein field 
equations are formulated in 
DW Hamil\-ton\-ian form 
as follows
\beqa    
\der_\al h^{\be\ga}
&=& 
\der {\cal H}  / \der Q^\al_{\be\ga} , \\ 
\der_\al Q^\al_{\be\ga}
&=& 
- \der {\cal H}   / \der h^{\be\ga}  ,  
\eeqa  
where eq.~(3.10) is equivalent  to the 
standard 
expression of the Christoffel symbols in terms of the metric 
 and  eq.~(3.11)  
 yields 
the vacuum  
Einstein equations 
in terms of  the Christoffel symbols. 

The present DW formulation originally was obtained in \cite{horava} 
using the theory of Lepagean equivalents. However, it can be 
derived also by straightforwardly applying the 
transformations leading to eqs. (1.1) 
to the Einstein truncated Lagrangian density: 
$$
{\cal L}_E = \frac{1}{16\pi G} h^{\mu\nu} (\Gamma^\lambda_{\mu\sigma}
\Gamma^\sigma_{\nu\lambda} -
\Gamma^\sigma_{\mu\nu} \Gamma^\lambda_{\sigma\lambda}).
$$

\subsubsection{Naive precanonical quantization. }  

Now, let us formally    
follow   the curved space-time version of precanonical quantization 
scheme and apply it to the above DW formulation of general relativity.  
This leads to the  following operator form 
of polymomenta $Q^\al_{\be\ga}$ 
\beq
\what{Q}{}^\al_{\be\ga} = -i\hbar \kappa 
\gamma^\alpha   
\left \{ \sqrt{g} 
\frac{\der}{\der h^{\beta\gamma}} \right \}_{ord}   
\eeq  
which is given  up to an  ordering ambiguity 
in the expression inside the curly brackets $\{ ... \}_{ord}$. 
By substituting this expression to (3.9) 
and performing a formal calculation using the assumption of 
the ``standard'' ordering 
(that the differential operators 
are all collected to the right) 
and  
relation  (3.6) for curved 
$\gamma$-matrices 
we obtain the operator form of the DW Hamiltonian density,  
also up to an  ordering ambiguity: 
\beq
\what{{\cal H}} 
= - 8\pi G\, 
\hbar^2\kappa^2 \frac{n-2}{n-1}  
\left \{  
 \sqrt{g} 
h^{\al\ga}h^{\be\de}\frac{\der}{\der h^{\al\be}} 
\frac{\der}{\der h^{\ga\de}} \right \}_{ord} 
+ (n-2)  \Lambda 
 \sqrt{g} ,  
\eeq  
where $\sqrt{g}$ can be obviously expressed in terms of 
our field variables $h^{\al\be}$. 


However, 
it should be pointed out that the above procedure 
of the construction of operators 
is rather of heuristic and formal nature. 
In fact, according to (3.8) classical polymomenta 
$Q^\al_{\beta\gamma}$ transform as 
 the 
connection coefficients 
while the operator associated with them in (3.12)  
is a tensor.   
Moreover, 
the classical DW Hamiltonian density (3.9) 
is an affine scalar density,  
while the operator constructed in (3.13)  
is a diffeomorphism scalar density. 
Therefore, we must clarify whether or not, 
or in which sense, the above procedure is meaningful.   
 
We shall argue below that expressions (3.12) and (3.13) are 
valid locally, i.e. in a vicinity of a point, 
while the information as to how to go from one space-time 
point to another, that is the structure of the connection, 
is given by the Schr\"odinger equation. 
This is very much 
along the lines of the precanonical  
approach to field quantization  
which can be viewed also as the ``ultra-Schr\"odinger'' 
picture, 
in which the space-time dependence is  
 totally transfered from operators to the wave function. 

\subsubsection{Covariant Schr\"odinger equation 
  for quantized gravity 
and the ``bootstrap condition.'' }

In order to understand the meaning of  
the specific realization of operators in Section 3.2.2 
let us remind first that  
the 
prescriptions of canonical  quantization  
are actually applicable only in a specific 
coordinate system and in principle 
require a subsequent ``covariantization.'' 
Second, let us note  that the consistency of 
the 
expressions (3.12) and (3.13)  
with the classical transformation laws 
could be achieved by 
adding an auxiliary  term in (3.12) which transforms as a connection.  
Then, the expression of the 
Christoffel symbols in terms of the polymomenta 
$Q^\al_{\be\ga}$ (c. f. eq. (3.8))    
\beq
\Gamma^\al_{\be\ga}= 8\pi G 
\left ( 
\frac{2}{n-1} 
\delta^\al_{(\beta } Q^\delta_{\ga)\delta} 
- Q^\al_{\be\ga} 
\right ) 
\eeq 
would yield an 
operator form of the Christoffel symbols 
\beq  
\what{\Gamma}{}^\al_{\be\ga} = 
- 8\pi i G\hbar\kappa \left \{ \sqrt{g} 
\left 
(\frac{2}{n-1} \delta^\al_{(\beta } \gamma^\sigma 
\frac{\der}{\der h^{\ga ) \sigma }} 
- \ga^\al \frac{\der}{\der h^{\beta \ga}} 
\right ) \right \}_{ord} 
+ \tilde{\Gamma}{}^\al_{\be\ga}(x) ,  
\eeq 
where the auxiliary (reference) connection 
is denoted  $\tilde{\Gamma}{}^\al_{\be\ga}(x)$. 
However, it is obvious that no arbitrary quantity like  
$\tilde{\Gamma}{}^\al_{\be\ga}(x)$ should be present in 
a desired background independent formulation. 

On the other hand, we can notice that our 
precanonically quantized operators arise essentially from 
the ``equal-point'' commutation relations  
 (c.f. eqs.~(2.3))  
and thus can be viewed as 
 locally 
defined  ``in a point''.  
In an infinitesimal vicinity 
of a point $x$ we always can chose 
a local reference system 
in which the 
auxiliary connection 
$\tilde{\Gamma}{}^\al_{\be\ga} (x) $ 
vanishes. 
Then one can assume that this is the 
reference system in which the expression 
(3.12) for operators $\what{Q}{}^\al_{\be\ga}$ 
is valid. 
However, when consistently implemented, 
this idea requires a subsequent "patching together" 
procedure in order to specify how and in which sense the 
operators determined in different points are 
related to each other. 
This procedure is likely to lead to extra terms in our 
generalized Schr\"odinger 
equation (3.7) 
because of 
the connection involved in the "patching together".  
In fact, in accord with the essence of the ``ultra-Schr\"odinger'' 
picture adopted here, when all the space-time dependence 
is transfered from operators to the wave function, 
it is natural to assume that the information about 
the "patching together", i.e. about 
passing from one space-time point into another,  
is actually controlled by the wave function 
and the Schr\"odinger equation it fulfils. 

This idea can be implemented as follows. 
At first we formulate a generalized 
Schr\"odinger equation (3.7) in the local 
coordinate system in the vicinity of a point $x$ 
in which the reference connection vanishes: 
$\tilde{\Gamma}^\al_{\be\ga} |_x$ = 0,  
and then covariantize the resulting equation 
in the simplest way. 
The first step leads to a locally valid equation 
 (c. f. (3.7)) 
\beq
i\hbar\kappa \sqrt{g}\gamma^\mu 
(\der_\mu + \hat{\theta}_\mu)  
\Psi = 
\what{{\cal H}} \Psi ,  
\eeq
where the local operator form of the coefficients of the spin  
connection $\what{\theta}{}_\mu$ (in the vicinity of $x$), 
as it follows from (3.4) and (3.15),   
is  given by 
\beqa 
\what{\theta}^A{}_{B\mu} 
\!&=&\!  
-8\pi i G \hbar \kappa 
\left \{ e^A_\al e^\nu_B \sqrt{g} 
\left ( 
\frac{2}{n-1} \delta^\al_{(\mu } 
\gamma^\sigma 
\frac{\der}{\der h^{\nu)\sigma }} 
- \ga^\al \frac{\der}{\der h^{\mu \nu}} 
\right ) \right \}_{ord} 
+\tilde{\theta}^A{}_{B\mu} |_x \nn \\
\!&=:&\! ({\theta}^A{}_{B\mu})^{op} + 
\tilde{\theta}^A{}_{B\mu} |_x , 
\eeqa  
where $({\theta}^A{}_{B\mu})^{op}$ 
 denotes  the first  (ordering dependent)  
operator term 
and $\tilde{\theta}^A{}_{B\mu}|_x$ denotes  
a reference spin connection which ensures the correct 
transformation law of (3.17). 
Note that in general 
$\tilde{\theta}^A{}_{B\mu}|_x \neq 0 $  
even if 
$\tilde{\Gamma}^\al_{\be\ga}|_x =0$.

Now, in order to formulate a generally covariant version of 
(3.16) we  notice that 
vielbeins do not 
enter the DW Hamiltonian 
formulation of general relativity 
on which the quantization in question is based. 
Therefore, within the present consideration 
they may (and can only) be 
treated as non-quantized classical 
$x$-dependent quantities: $e^\mu_A = \tilde{e}{}^\mu_A(x)$.  
%
On another hand, 
the  bilinear combination  of vielbeins 
$e_A^\mu e_B^\nu\eta^{AB}$ is 
the metric tensor 
 $g^{\mu\nu}$ 
which {\em is} a 
variable  quantized 
(in the ``ultra-Schr\"odinger''  picture used here)
as an $x$-independent quantity.   

Both aspects  
 can be 
 reconciled  
in agreement with the correspondence principle 
by requiring 
 the bilinear combination of vielbeins 
 to be 
consistent with the mean  value  of the metric, 
$\left < g^{\mu\nu}\right >(x)$, 
i.e. 
\beq  
\tilde{e}{}^\mu_A(x)
\tilde{e}{}^\nu_B(x) \eta^{AB} = 
\left < g^{\mu\nu}\right >(x) ,      
\eeq   
 where the latter 
is given by 
averaging  over the space of 
 the 
metric components by means of 
the wave function $\Psi(g^{\mu\nu},x^\mu)$ \, (c.f. eq. (2.8)):     
\beq 
\left <g^{\mu\nu}\right >(x) = 
\int [d g^{\al\be}] 
~\Psib (g,x) g^{\mu\nu} \Psi(g,x)  ,   
\eeq 
with 
 the 
invariant integration measure 
 given by 
(c.f. \cite{misner})\footnote{To avoid a possible confusion let us 
notice that the scalar product in (3.19) and the 
finite dimensional diiffeomorphism invariant 
integration measure (3.20) are mathematically well defined, 
 in contrast to their infinite dimensional 
counterparts in quantum geometrodynamics based on 
the Wheeler-DeWitt equation. } 
\beq
[d g^{\al\be}] = 
\sqrt{g}^{\,(n+1)} \prod_{\al\leq \be} d g^{\alpha\beta } . 
\eeq  
Hence,  the vielbein field  $\tilde{e}{}^\mu_A(x)$   
is set to be  
determined by the consistency  with 
the averaged metric field.  
In doing so the local orientation of 
vielbeins is still arbitrary 
but it can be fixed by a proper coordinate 
 (gauge) condition on  
 an average 
vielbein field $\tilde{e}{}^\mu_A(x)$. 
A natural idea is to use the 
averaged vielbein field $\tilde{e}{}^\mu_A(x)$ to specify 
the quantities in the covariantized version of (3.16), 
like the reference spin connection, for which no operator 
expression  can be found  within the metric formulation.  


Now, a diffeomorphism covariant 
version of (3.16)   
can be written 
in the form 
\beq  
i\hbar\kappa 
\tilde{e} 
\tilde{e}^\mu_A(x)\ga^A(\der_\mu + 
\tilde{\theta}_\mu (x)) \Psi  
+ 		
i\hbar\kappa (\sqrt{g} \gamma^\mu {\theta_\mu})^{op} \Psi 
=  \what{{\cal H}}\Psi  
\eeq 
which involves the 
self-consistent average vielbein field $\tilde{e}^\mu_A(x)$ 
given by the ``bootstrap condition'' (3.18), (3.19)  
and the corresponding spin connection $\tilde{\theta}_\mu (x)$.    
This makes the equation essentially nonlinear and integro-differential.   
However,  the corresponding ``non-locality''  
is totally confined to the inner space of the metric components, 
over which the integration is implied in (3.19),  
and, therefore,  does not alter the locally causal character 
of the equation in (a self-consistent, averaged) space-time. 
At the same time the nonlinearity in the left 
hand side of (3.21) specifies the averaged space-time described 
by the tilded quantities and 
does not alter the quantum dynamics in the inner space which 
is governed by the linear operator $\what{{\cal H}}$, eq.~(3.13), 
and, therefore, 
is consistent with the superposition principle. 
Moreover, inasmuch as the tilded quantities 
present in eq. (3.21) 
 are introduced 
 as resulting from the quantum averaging  
self-consistent with the underlying quantum dynamics 
of the wave function,  
they represent not an arbitrary a priori fixed classical background but  
an averaged self-consistent 
space-time geometry which enables us to formulate the Dirac-like 
equation for the wave function, as it is characteristic to the precanonical 
quantization approach. In this sense the formulation can be viewed 
as background independent.  
  
The explicit form of the operator part of the spinor connection term 
in (3.21), $(\sqrt{g} \gamma^\mu {\theta_\mu})^{op}$,    
can be derived  from (3.17).  
By assuming the ``standard'' ordering  of operators 
in an intermediate calculation 
and replacing, when appropriate, the 
 appearing 
therewith 
bilinear combinations   
of vielbeins with the metric tensor, 
we obtain 
\beq
(\sqrt{g} \gamma^\mu {\theta_\mu})^{op} 
= - n 
\pi i G \hbar\kappa 
\left \{ \sqrt{g} h^{\mu\nu} 
\frac{\der}{\der  h^{\mu\nu}} \right \}_{ord} . 
\eeq 

The $x$-dependent reference spin connection 
term $\tilde{\theta}_\mu (x)$ 
in (3.21) 
is related to 
the 
average self-consistent vielbein field 
$\tilde{e}{}^A_\mu (x)$,  
given by 
  the ``bootstrap condition'' 
(3.18), (3.19)  and a proper coordinate condition,   
by the classical expression 
\beq
\tilde{\theta}{}_\mu^{AB} (x) =  
\tilde{e}{}^{\al [A} 
\left 
(2 \der_{[\mu} \tilde{e}{}^{B ]}{}_{\al ]} 
+ \tilde{e}{}^{ B] \be} \tilde{e}{}^C_\mu \der_\be 
\tilde{e}{}_{C \al}  
\right )
\eeq
which is equivalent to (3.4).   

Lastly, let us note that in order  
to distinguish a physically relevant information 
in (3.21)  we need to impose a gauge-type condition 
on $\Psi$. The meaning of this condition is to  
single out a specific wave function $\Psi (h^{\mu\nu},x^\al)$ 
from the class of wave functions which lead to the 
averaged metric fields which are ``physically equivalent''. 
For example, if one is to impose the 
De Donder-Fock harmonic gauge on the averaged metric 
field then the corresponding condition on the wave function 
reads: 
\beq 
 \der_\mu (\left < h^{\mu\nu} \right >(x)) = 0 ,   
\eeq 
where $\left < h^{\mu\nu} \right >(x)$ is given 
similarly to (3.19) and (3.20). 

Thus, we conclude that within 
precanonical quantization based on DW formulation 
the 
quantized gravity 
is 
described by a generally covariant  
generalized Schr\"odinger 
equation (3.21),  
with  the operators $\what{{\cal H}}$ and 
$(\sqrt{g} \gamma^\mu {\theta_\mu})^{op} $ given 
respectively by 
 (3.13) and  (3.22),  
and 
the 
supplementary 
``bootstrap condition'' (3.18) which 
specifies the tilded quantities  
representing the self-consistent average space-time 
geometry. 

The solutions of these equations 
$\Psi(g^{\mu\nu}, x^\al)$ 
can be interpreted as the probability amplitudes  
 of finding 
the values of the components 
of the metric tensor in the interval 
$[g^{\mu\nu}$ -- $(g^{\mu\nu} \!+\! dg^{\mu\nu}) ]$ 
in an infinitesimal vicinity of the point $x^\al$. 
Obviously, this description is very different from 
the conventional quantum field theoretic one and 
its physical  
significance remains to be explored. 
 It is interesting to note, 
however, that it opens an intriguing 
possibility to 
approximate 
the ``wave function of the Universe'' 
by the fundamental solution of  equation (3.21).  
This solution is 
expected to 
describe  
 an expansion 
of the wave function from the 
primary    
 ``probability lump'' 
of the Planck scale   
and assigns a meaning to the ``genesis of the space-time'' 
in the sense that 
 the observation of 
the space-time points 
beyond the primary 
``lump'' becomes more and more  
 probable with the spreading  of the wave function.    
The self-consistency encoded in the ``bootstrap condition''  
 obviously plays a crucial role in 
 this process: in a sense, the 
wave function itself determines, or ``lays down'', 
the space-time geometry it is to propagate on.    

\section{ 
 Concluding remarks  
}

The problem of quantization of gravity has been treated here 
from the point of view of 
precanonical quantization based on the structures 
of the De Donder-Weyl theory viewed as  
a manifestly covariant generalization of 
the Hamiltonian formulation 
from mechanics to field theory.

The De Donder-Weyl Hamiltonian formulation is an attractive 
starting point for quantization of gravity as it 
does not  distinguish  between 
space and time dimensions  
and represents the fields essentially as systems 
{\em varying } in space-time  
rather than as infinite dimensional systems {\em evolving } in time. 
The De Donder-Weyl Hamiltonian equations (1.1), 
which are equivalent  to the Euler-Lagrange equations, 
 provide us with the Hamiltonian-like  description of 
 this type of varying in space-and-time. 
These equations are formulated using the finite dimensional analogues 
of the configuration space - the space of field and space-time 
variables, and the phase space - the space of field and space-time 
variables and polymomenta.  

The quantum counterpart of the theory is formulated also 
on 
a 
finite dimensional configuration 
space of field and space-time variables. 
The corresponding wave { function} $\Psi(x^\mu, y^a)$ 
is naturally interpreted as 
the 
probability amplitude of 
a 
field 
to take a value  
 in  
the interval [$y - (y\!+\!dy)$] in the vicinity of 
the space-time point $x$. 
In doing so all the dependence on 
 a 
space-time location 
is transfered 
from operators to the wave function, 
  corresponding to  
 what we called the 
 ``ultra-Schr\"odinger'' 
 picture.   

It should be noted that despite the finite dimensionality 
of the constructions of the precanonical approach no 
field degrees of freedom, understood in the conventional sense, 
are ignored. This is evident, on the classical level, 
from the fact that 
the DW Hamiltonian equations  
are equivalent to the  field equations and, 
on the quantum level, from the observation that 
 our generalized Schr\"odinger equation, eq. (2.6),  
reproduces the field equations in the 
classical limit and 
also can be related to the standard 
functional differential  Schr\"odinger equation  \cite{ik-schr}. 

 It is clear that the foundations of the present approach 
to field quantization are very different from those of 
the standard quantum field theory. Because of this 
conceptual distance it is not easy to establish a connection 
between the both. Unfortunately, 
a poor understanding of this issue so far 
has been hindering specific applications of the approach  
(see, however, \cite{castro} for a recent attempt to apply it 
to quantization of $p$-branes).  
  
Nevertheless,  
 the already understood character of 
connections   
between the De Donder-Weyl theory 
and the standard Hamiltonian formalism 
\cite{gotaymulti1,sternberg} 
seems to provide us with a clue 
  to a possible approach to 
this problem.         
In fact,   
the standard symplectic form  
and the standard equal-time canonical brackets  
 in field theory 
can be obtained by   
 integrating the polysymplectic form $\Omega$ 
 and the canonical brackets (2.3) 
 over  the Cauchy data surface $\Sigma: (y^a=y^a(\bx), t=const)$ 
in the covariant configuration space $(y^a,x^\mu)$ \cite{romp98}. 
Similarly,  
the standard functional differential 
field theoretic 
Hamilton and Hamilton-Jacobi equations 
  can be deduced from the 
partial differential DW Hamiltonian and     
the DW Hamilton-Jacobi        
equations 
by  restricting  
the quantities of the DW formulation to a Cauchy data 
 surface 
 $\Sigma$ 
 and then integrating over it. 
 It is natural to  expect that a similar connection 
 can be  established  between the 
the elements of the precanonical approach 
to field quantization 
and those 
of the standard canonical quantization. 

 A related way to find a connection with 
the conventional quantum field theory is to view the 
Schr\"odinger wave functional $\Psi([y(\bx)],t)$ 
(see e.g. \cite{hatfield})  
as a composition of amplitudes given by our 
wave function $\Psi(y,\bx,t)$ restricted to $\Sigma$.  
In fact, developing an earlier demonstration of this connection 
in the ultra-local approximation \cite{qs96,lodz98} we have 
shown recently \cite{ik-schr} that the Schr\"odinger wave functional  
can be represented as the trace of the positive frequency part of the 
continual product over all spatial points 
of the values of the  wave function $\Psi(y^a,x^\mu)$  
restricted to a Cauchy surface. Besides, it has been 
shown that using this ansatz 
the standard functional differential Schr\"odinger 
equation can be  derived from our Dirac-like generalized 
Schr\"odinger equation, eq. (2.6). It is natural to ask if this kind 
of interplay between precanonical and canonical quantization 
could be extended to gravity in order 
to understand a possible relation between 
the Dirac-like wave equation for quantized gravity proposed in 
Sect. 3.2.3 and the Wheeler-DeWitt equation.


 The character of the 
inter-relations,  as outlined above, 
between the DW formulation 
 (and more general Lepagean theories \cite{lepage,dedecker,krupka}) 
 and the conventional  
 canonical 
 formalism 
is the reason to refer to the former as 
 the  {\em precanonical }  formalism. 
The term reflects 
 an intermediate position of the DW formulation 
(and its Lepagean generalizations)   
between the covariant Lagrangian and the 
 ``instantaneous''  
  Hamiltonian  levels of description.  
Note that in mechanics, $n=1,$ 
 the 
 precanonical 
 description coincides with 
 the 
 canonical  
 one and it is only in field theory, $n>1,$ they become different. 
The same is valid for {\em  precanonical quantization} 
underlying the present approach to quantization of  gravity. 

It should be noticed that the application of the precanonical  
framework to gravity immediately 
rises many questions to which 
no final answers can be given 
 as 
yet. 
Some of these, such as, for example, 
  
(i) how the spinor wave function 
is reconciled with the boson vs. fermion nature of the 
fields we quantize;    
 
(ii) if it can or should be replaced 
by a more general Clifford algebra valued wave function;   
 
(iii) to which extent  one can rely on 
the prescription of quantum 
averaging (2.8) if the underlying scalar product 
is neither positive definite nor $x$-independent;   

(iv) how to quantize the operators more general than those  
entering the precanonical brackets (2.3);   
and, at last,  

(v) how to calculate the observable quantities of interest 
 in field theory using the precanonical framework, \\
concern rather 
the precanonical approach in general and are still 
being investigated. We hope to address them elsewhere.   
Let us  instead concentrate here on a few questions related to 
the specific application of precanonical quantization 
to  general relativity.

A severe difficulty   we encountered 
is related to 
the non-tensorial nature of 
 the basic quantities 
(the polymomenta and the DW Hamiltonian)
of the DW formulation of general relativity  
in Section 3.2.1, 
which is in disagreement with  
the tensorial character of operators 
which only can be 
constructed  
 (in a background independent fashion) 
as their quantum counterparts.  
The origin of this difficulty is in the fact that the 
DW formulation of general relativity in Section 3.2.1 
is based  essentially on the Einstein 
non-covariant truncated Lagrangian density,    
which contains no second-order derivatives of the metric, 
instead of the generally covariant 
Einstein-Hilbert Lagrangian density $\sqrt{q} R$.   
 To use the latter we would need a generalization of  
the precanonical constructions 
outlined in Section 2.1 and 2.2 to 
 the 
second order irregular Lagrangians 
(see, e.g., \cite{krupka,gotay-ext,azcarraga} 
and the references therein), 
which is largely not developed as yet.  
 The vielbein formulation of  general relativity in the 
second order formalism would face a similar difficulty. 
An attempt to use the first order (Palatini) 
formalism (c.f. \cite{gimmsy,esposito})  
 also leads to highly irregular Lagrangians 
 which  require a proper adaptation of the 
 precanonical treatment, yet to be 
 developed.  

In the approach of the present paper 
    the difficulty mentioned above is circumvented 
 by quantizing locally, in a vicinity of a point, and then covariantizing.  
 Though this procedure 
 involves  
external elements,  
 such as 
a reference 
vielbein field 
$\tilde{e}{}^\mu_a(x)$ 
and the corresponding spin connection  
$\tilde{\theta}_\mu (x)$,      
 which enter  
into the generalized 
Schr\"odinger equation (3.21) 
as non-quantized quantities,    
  those 
are not   arbitrary since the correspondence principle 
requires 
the reference vielbein field  
to be consistent with the mean value 
of  the metric. This requirement leads to 
a self-consistency in the theory in the form of 
the ``bootstrap condition'' (3.18), (3.19)  
 which 
connects the 
bilinear combination of 
 the reference 
vielbein fields  
$\tilde{e}{}^\mu_a(x)$ 
with the quantum 
  mean value of the metric. 
By this means the allowable 
 classical geometry 
 emerges in the theory  
as an approximate 
notion - a result of quantum averaging - 
 in a 
 self-consistent with the underlying quantum dynamics  
way. In this sense the theory 
 appears to be 
 independent of an arbitrarily chosen background.   

 This point of view, though  looking  less radical than 
the usual denial of any background geometrical structure 
in quantum theory of gravity,  
 which  
 is known to lead to 
the 
most of the conceptual difficulties 
 of the latter,  
 seems to offer a working alternative 
 to the currently more popular attempts to proceed  from a 
  specific model of quantum ``pregeometry'' near to the Planck scale,  
 be it a discrete space-time, 
 a space-time foam, 
 a non-commutative or fuzzy space-time,  
 or 
 the spin networks and a spin foam 
 recently proposed within the Ashtekar program 
(for a review and further references see, e.g.,  \cite{rovelli}). 

  On another hand, the appearance 
  in the left hand side of (3.21) 
  of, essentially, an averaged Dirac operator,  
may imply an approximate,  ``smea\-red down,''   
not ultimately quantum, character of the description achieved 
here. 
In this case a further step could be required  which would allow us 
to treat the Dirac operator in the  left hand side of (3.7) 
beyond the framework of classical  
geometry.   
 In this case  
a proper insight into a quantum pregeometry 
could be important indeed.

 Let us mention also   
 that the coefficients involving $n$ in (3.13) and (3.22) 
at the present stage cannot be considered   
as reliably established.   
 This is related both to the ordering   
ambiguity and to the unreliability of the results obtained by 
formal substitution of polymomenta operators (3.12) 
to classical expressions ({for example,  applying 
the similar procedure to the DW Hamiltonian of a  massless 
scalar field $\phi$ yields the operator 
$-\frac{n}{2} \hbar^2 \kappa^2 \der^2_{\phi\phi}$ 
instead of the correct one 
$-\frac{1}{2} \hbar^2 \kappa^2 \der^2_{\phi\phi}$} \cite{qs96,bial97}).  
 Note also, that at this stage it is also 
 rather 
difficult to choose  
between the formulation based on the operator 
of DW Hamiltonian $\what{H}$ 
and  that based on the corresponding density 
$\what{{\cal H}}$. In the 
 former  case,  
the generalized Schr\"odinger 
equation, eq.  (3.7),  would be  modified as follows: 
$i\hbar\kappa 
\what{\mbox{$ \gamma^\mu \nabla_\mu$}} \Psi = 
\what{{H}} \Psi$,  
which  in general is different from (3.7)  
due to the ordering ambiguity. 
A preliminary consideration of 
 the 
 toy one-dimensional models corresponding to the formulations using  
$\what{H}$ and $\what{{\cal H}}$ 
 respectively 
indicates \cite{inprogress}  
that the latter formulation, which leads to a toy model 
similar 
to that discussed long ago by Klauder \cite{klauder}, 
 seem to reveal 
more interesting 
behaviour 
and thus might be more suitable.  
However,  to present more conclusive results, 
an additional analysis, 
possibly based on quantization of more general 
dynamical variables than those involved in the 
 precanonical brackets  (2.3), is required. 
 Besides, as we have already 
pointed out,   
the vielbein  formulation of general relativity  
    can be more adequate to the application of precanonical quantization 
 to gravity, though it is unlikely to be a panacea  from the 
problems we have outlined above. 
 The corresponding  analysis is in progress and we hope 
to   report  on the results elsewhere. 

In conclusion, let us summarise in short the potential 
advantages  
of the present approach. 
The obvious advantage is its manifest covariance (more precisely, the 
starting point and the resulting equations  are covariant though 
some intermediate steps 
still are not).  
This allows us to 
avoid 
the usual restriction to globally hyperbolic 
space-times which is necessarily imposed in canonical quantum gravity.    
However, this advantage, 
though potentially important for considering the 
expected quantum topology and signature changes 
in quantum gravity, still could be viewed by a sceptic 
as a purely technical achievement.    
Another technical advantage is that the analogue of the 
Schr\"odinger  equation and other elements of the 
formalism 
reveal no problems with their mathematical definition 
(the ordering problem encountered here is, in fact,  
not more complicated  than that in quantum mechanics),   
 in contrast to the approaches based on  the Wheeler-DeWitt equation or 
the path integral.  
This advantage, however, is 
in-built in the precanonical approach itself, 
which avoids treating fields as infinite-dimensional systems, 
and is not specific to the quantum gravity.  

As far as the physical aspects of the theory are concerned,    
an  intriguing feature of the approach 
 is the appearance  of a self-consistently 
 incorporated 
  averaged 
  vielbein field 
in the generalized Schr\"odinger   
equation (3.21).  
This enables us to avoid 
the direct tackling with the 
problems of quantum pregeometry, i.e. 
an ultimate 
description of ``quantum space-time'' near to the Planck scale  
(for a recent discussion see  \cite{isham,isham99b}),      
 which are usually viewed to be the central issue of 
 quantum gravity. 
Nevertheless, 
in spite of not giving an insight  
as to what the 
quantum space-time, or pregeometry, could be 
 the present treatment  refers to the classical space-time  
only as an approximate notion  
resulting from  the quantum averaging  
 and  
 a 
self-consistency. No arbitrarily fixed 
classical background geometry is 
been involved. This essentially amounts  
to a background independence of the formulation. 

Besides, 
the appearance of a  self-consistent vielbein field 
provides us with a 
framework for discussing the problem of emergence of 
classical space-time  in quantum gravity 
(for a recent review see  \cite{isham99a}).  
Moreover, 
it could shed light on the problem of interpretation  
(or, that    
of an 
``external observer'')  
in quantum cosmology   
 since 
 the generalized Schr\"odinger equation (3.21) 
 essentially describes a sort of self-referential quantum system, 
 in the sense that the 
 self-consistent averaged vielbein field  
 can be viewed as 
 the vielbein field 
representing 
 the macroscopic ``self-observing'' 
 degrees of freedom of a quantum gravitational system.


    \section*{Acknowledgments} 

I thank J. Klauder  for drawing my 
attention to his earlier papers \cite{klauder},  
A. Borowiec and M. Kalinowski for  valuable remarks,       
C. Castro and J. S\l awianowski 
for stimulating encouragement and useful comments, 
and M. Pietrzyk for her effective numerical analysis 
of a simplified 1-D version of eq. (3.21) 
which was very helpful for understanding of its behavior.  
I am also grateful to 
A. Wipf\, and the Institute of 
Theoretical Physics of the Friedrich Schiller 
University of Jena for their kind hospitality 
and excellent working conditions 
which enabled me to write this paper.




\end{document}